\global\let\ifmypprint\iffalse 
\def\mypprint{\global\let\ifmypprint\iftrue}
\global\let\iftorefs\iffalse
\def\torefs{\global\let\iftorefs\iftrue}
\global\let\dofloatfig\iffalse
\def\floatthefig{\let\dofloatfig\iftrue}
\mypprint 
\ifmypprint
\documentstyle[twocolumn,prl,aps]{revtex}
\floatthefig
\else
  \documentstyle[preprint,aps]{revtex}
  \iftorefs 
    \catcode`\@=11 
    
    \def\figure{\let\@capwidth\columnwidth\@float{figure}}
    \let\endfigure\end@float
    \@namedef{figure*}{\let\@capwidth\textwidth\@dblfloat{figure}}
    \@namedef{endfigure*}{\end@dblfloat}
    \catcode`\@=12 
     \floatthefig 
  \fi
\fi

\input epsf.tex
\begin{document} 
\draft
\title{Hexagons, Kinks, and Disorder in Oscillated Granular Layers} 
\author{Francisco Melo \cite{fmelo} }
\address{Departamento de F\'{i}sica, Universidad de Santiago,\\
	 Avenida Ecuador 3493, Casilla 307 Correo 2, Santiago, Chile}
\author{Paul B. Umbanhowar\cite{pbu} and Harry L. 
Swinney\cite{hls} }
\address{Center for Nonlinear Dynamics and Department of 
Physics,\\
         The University of Texas at Austin,
         Austin, TX 78712}
\date{\today}
\maketitle

\ifmypprint
  \widetext
\fi

\begin{abstract} Experiments on vertically oscillated granular layers
in an evacuated container reveal a sequence of well-defined pattern
bifurcations as the container acceleration is increased.  Period
doublings of the layer center of mass motion and a parametric wave
instability interact to produce hexagons and more complicated patterns
composed of distinct spatial domains of different relative phase
separated by kinks (phase discontinuities).  Above a critical
acceleration, the layer becomes disordered in both space and time.
\end{abstract}


\ifmypprint
  \narrowtext 
\fi


The transport, mixing, and segregation of granular materials is
important in industries ranging from food to mineral processing, yet a
basic understanding of the physical mechanisms underlying the
collective dynamics of grains is lacking.  Recent experiments on
vertically vibrated granular materials show a variety of phenomena
including heap formation and convection
\cite{raj,laroche}, size segregation \cite{nagel1}, and traveling
waves \cite{behringer}.  These phenomena, although of practical
importance, are caused by surrounding gas and/or sidewall driving
\cite{pak}, leaving open the question of whether there is any 
intrinsic self-organized behavior in these systems.  We report here
robust patterns that arise spontaneously, not from interstitial gas or
sidewall forcing, but from correlations induced by multiple collisions
between the grains and by the coherent motion of the particle layer
and the container.


Our experiments on vertically vibrated granular layers yield spatial
patterns composed of standing waves that oscillate at either one-half
or one-quarter of the drive frequency $f$.  Spatial domains of
different relative phase separated by phase discontinuities (kinks)
appear in all patterns except those just beyond the initial
instability.  Examples of stripes, squares, hexagons, kinks, and a
disordered state are presented in Fig.\ \ref{pics}.  Figure
\ref{phasd} shows the stability region for each pattern as a function
of $f$ and the dimensionless acceleration amplitude $\Gamma =
4\pi^2f^2A/g$, where $2A$ is the peak-to-peak amplitude of the
sinusoidal displacement of the container and $g$ is the gravitational
acceleration.  The transitions are well-defined, only weakly dependent
on $f$, and non-hysteretic (except for the transitions to squares).
We will now describe the experimental methods and then show how the
patterns and their associated thresholds arise from the interaction of
two distinct mechanisms: a parametric wave instability
\cite{douady,melo} and period doublings.

\dofloatfig
\pagebreak
\vspace*{1.2in}
\noindent
\begin{figure}
\epsfxsize=8.5truecm
\centerline{\epsffile{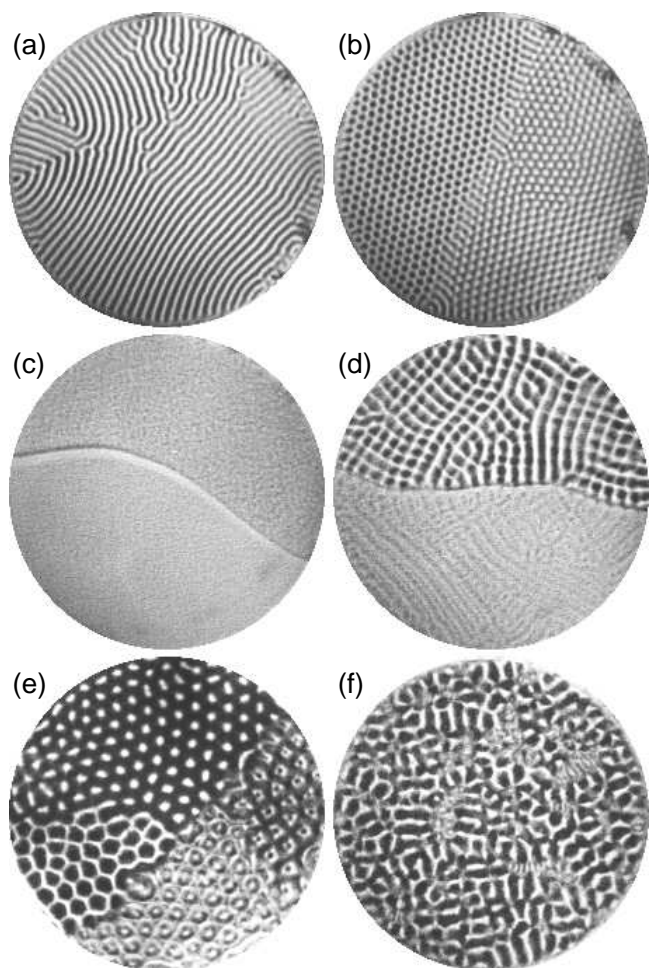}}
\smallskip
\caption{Patterns in a 1.2 mm deep layer at $f$ = 67 Hz: (a) $f/2$
stripes ($\Gamma$ = 3.3), (b) $f/2$ hexagons ($\Gamma$ = 4.0), (c)
flat with kink ($\Gamma$ = 5.8), (d) competing $f/4$ squares and
stripes ($\Gamma$ = 6.0), (e) $f/4$ hexagons ($\Gamma$ = 7.4), and (f)
disorder ($\Gamma$ = 8.5).}
\label{pics}
\end{figure}
\fi

\dofloatfig
\begin{figure}
\epsfxsize=8.5truecm
\centerline{\epsffile{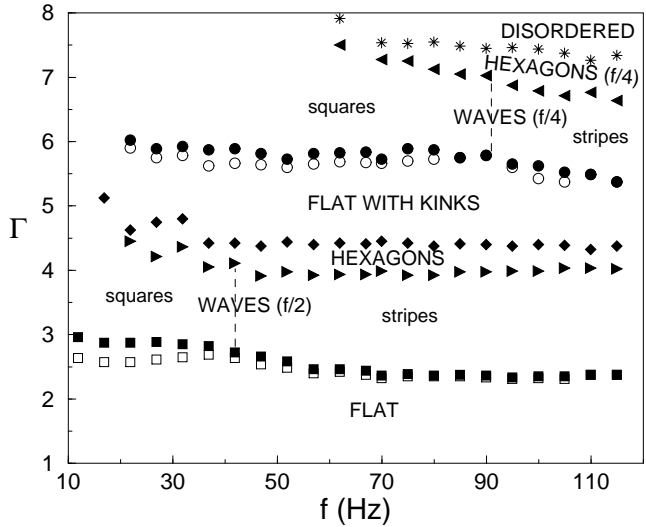}}
\smallskip
\caption{Stability diagram showing transitions in a 1.2 mm deep layer.
The vertical dashed lines indicate the frequencies above which only
stripes appear in the wave regimes.  Closed (open) square and circular
symbols denote the transitions to waves with increasing (decreasing)
$\Gamma$.}
\label{phasd} 
\end{figure}
\fi


A layer of \mbox{0.15--0.18 mm} diameter bronze spheres is placed in
the bottom of a cylindrical container that has inner diameter
\mbox{127 mm} and height \mbox{90 mm}; the wall and lid are Lucite
while the base is aluminum to reduce electrostatic effects.  The layer
is 7 particles deep except for the data in Fig.\ \ref{ftime}, where a
thicker layer (12 particles deep) improves the signal-to-noise ratio
for the acceleration measurements.  The container is evacuated to 0.1
Torr, a value at which the volumetric effects of the gas are
negligible \cite{pak} and heaping is not observed.  An
electromechanical vibration exciter drives the container and the
resulting acceleration is measured to a resolution of $0.01 g$.
Patterns are illuminated at low angle with a strobe light and are
recorded with a video camera located above the container.


The instabilities leading to the different patterns can be understood
in terms of two dimensionless parameters that characterize the
dynamics of the layer: $\tau = f t_{{\rm flt}}$, the layer free-flight
time, and $\gamma = v_c/g t_{{\rm flt}}$, which is approximately the
acceleration of the granular layer relative to the plate during the
time of collision ($v_c$ is the relative collision velocity)
\cite{gamma}.  We directly measure $\tau$, while we deduce $\gamma$
from a simple one-dimensional (1D) model of an inelastic ball on an
oscillating plate.  Figures \ref{ftime}(a) and \ref{ftime}(b) show the
dependence of $\tau$ and $\gamma$ on the driving acceleration
$\Gamma$, and Table
\ref{table1} indicates how the patterns can be characterized by the
values of $\tau$ and $\gamma / \gamma_{{\rm crit}}$, where
$\gamma_{{\rm crit}}$ marks the onset of parametric waves.  


In our model system, the ball is completely inelastic because the
collisions of the particle layer and container are almost completely
inelastic due to multiple internal grain collisions.  The ball motion
is computed \cite{mehta} by assuming that free-flights begin whenever
the ball and plate  
\dofloatfig
\begin{figure}
\epsfxsize=8.5truecm
\centerline{\epsffile{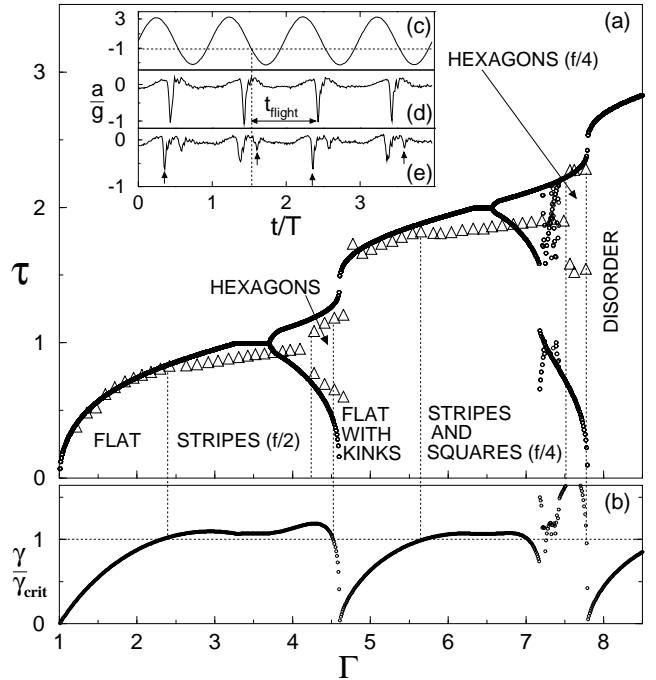}}
\smallskip
\caption{Comparison of observations for a particle layer to
calculations for a model consisting of a 1D completely inelastic ball.
(a) Flight times for the layer (1.9 mm deep, $f = 67$ Hz), $\tau_{{\rm
expt}}$ ($\bigtriangleup$), and for a completely inelastic ball,
$\tau_{{\rm calc}}$ ($\circ$).  (b) The dimensionless layer collision
acceleration $\gamma / \gamma_{{\rm crit}}$ (the average value is
displayed where $\gamma$ is multivalued).  (c) Container acceleration
(collisions removed). (d) and (e) Impact acceleration for stripes
($\Gamma=3.3$) and hexagons ($\Gamma = 4.2$) respectively (vertical
arrows indicate successive collisions of regions with the same
relative phase).}
\label{ftime} 
\end{figure}
\noindent 
\fi
are in contact and the plate acceleration $a(t) <
-g$.  Figure \ref{ftime}(a) compares the calculated flight time
$\tau_{{\rm calc}}$ with the measured flight time $\tau_{{\rm expt}}$
as a function of $\Gamma$. (Figures \ref{ftime}(c) and \ref{ftime}(d)
indicate how $\tau_{{\rm expt}}$ is measured.) In the absence of
parametric waves, the layer and ball motions are nearly the same:
$\tau_{{\rm expt}} = \tau_{{\rm calc}}$.  When parametric waves are
present the layer is dilated at take-off, which reduces the effective
take-off velocity of the layer and consequently decreases $\tau_{{\rm
expt}}$.  Since $\gamma$ is calculated from model values of $v_c$ and
$t_{{\rm flt}}$, it is only accurate when parametric waves are absent
($\gamma < \gamma_{{\rm crit}}$).  Because $v_c$ and $t_{{\rm flt}}$
are both proportional to $1/f$, $\gamma$ is independent of $f$; this
is consistent with Fig.\ \ref{phasd}, which shows that the parametric
wave transitions depend only weakly on $f$.


For $\Gamma > 1$, on each cycle the layer loses contact and later
collides with the container.  However, it is not until $\Gamma \approx
2.4$ that the parametric wave instability first occurs ($\gamma = 
\gamma_{{\rm crit}}$) and the flat
layer bifurcates to waves oscillating at $f/2$, as shown in Fig.\
\ref{pics}(a) \cite{ework}.  Vertical lines extending from Fig.\
\ref{ftime}(a) to Fig.\ \ref{ftime}(b) mark the experimental values of
$\Gamma$ where transitions associated with parametric waves occur.
The $f/2$ wave patterns are squares for \mbox{$f < 24$ Hz} and stripes
for \mbox{$f > 40$ Hz} (Fig.\ \ref{phasd}); at intermediate values of
$f$, both patterns compete.  For squares the difference between the
stability threshold for increasing and decreasing $\Gamma$ is
about twenty percent, while for stripes the hysteresis is
small and perhaps even zero.


Hexagonal patterns arise spontaneously from $f/2$ waves when the
vertical motion of the layer undergoes a period doubling bifurcation.
Period doubling occurs when $\tau_{{\rm expt}}$ exceeds unity and
subsequently becomes double valued (see Fig.\ \ref{ftime}(a)).  As
Fig.\ \ref{ftime}(e) shows, alternate collisions now occur before and
after $a_c = -g$; the layer is effectively forced at both $f$ and
$f/2$ because successive collisions occur with different $v_c$ and
$t_{{\rm flt}}$.  Like $f/2$ waves, hexagonal patterns return to the
same configuration after two periods but with an important difference:
time translation by one period is no longer equivalent to a spatial
shift of one-half wavelength.  Instead, two distinct patterns appear
in alternate cycles --- a set of isolated peaks on a triangular
lattice become, on the next oscillation, hexagonal cells, each one
centered on a former peak location.  The amplitude of the cellular
phase is maximum at the end of the long flight, while the amplitude of
the peaked phase is maximum at the end of the short flight.  Figure
\ref{phasd} shows that the $\Gamma$ value for transition to hexagons
($\Gamma \approx 3.9$) is nearly constant for frequencies above where
$f/2$ waves show a strong decrease in hysteresis.


Spatial kinks separating domains of different relative phase appear at
the period doubling bifurcation.  Kinks arise because the period
doubled motion is degenerate by $\pi$ in phase: a portion of the layer
can begin its motion on one cycle or the next.  For hexagons, this
degeneracy manifests itself in the simultaneous appearance of both
peaked and cellular spatial domains separated by a well-defined phase
defect, as shown in Fig.\ \ref{pics}(b).


As $\Gamma$ is increased further, the amplitude of the hexagonal
pattern decreases until $\gamma / \gamma_{{\rm crit}} < 1$ and
parametric waves disappear.  The layer now consists of flat domains
connected by a kink, as shown in Fig.\ \ref{pics}(c).  Physically,
this resonance corresponds to $v_c \approx 0$: the layer lands gently
on the plate.  Flat domains with kinks arise solely from the period
doubling instability and appear to be the 2D analogue of the 1D
subharmonic instability in a glass bead layer reported by Douady {\it
et al.\ } \cite{douady2}.

Beyond onset of the flat domains with kinks, $\tau_{{\rm expt}}$ is
double valued as for the hexagonal patterns, but as $\Gamma$ is
increased the time of the shorter flight goes to zero and $\tau_{{\rm
expt}}$ again becomes single valued.  In both cases the motion is
period doubled: in the former, flights are composed of long and short
jumps (effective two-frequency forcing), while in the latter, flights
can begin on one cycle or the next but the time between successive
collisions is always $2 / f$.  In this latter state, $\tau_{{\rm
expt}} \approx \tau_{{\rm calc}}$, indicating that the layer is
compact at take-off.


Parametric square and striped patterns reappear when again $\gamma / 
\gamma_{{\rm crit}} > 1$ at $\Gamma \approx 5.7$.  Because the time
between collisions is two periods, the waves oscillate at $f/4$ and,
as Fig.\ \ref{pics}(d) shows, appear in separate domains whose motions
differ in phase by $\pi$ (one domain is taking-off while the other is
in mid-flight).  Otherwise $f/4$ waves are the same as
$f/2$ waves: the transition is accompanied by a decrease in
$\tau_{{\rm expt}}$ with respect to $\tau_{{\rm calc}}$ and squares
appear subcritically at lower $f$, while at higher $f$ stripes occur
essentially without hysteresis.


Although the layer motion is period doubled for $f/4$ squares and
stripes, hexagons do not appear because $\tau_{{\rm expt}}$ is single
valued; the layer experiences the same conditions on consecutive
collisions.  However, when $\tau_{{\rm expt}} > 2$, the layer motion
undergoes a second period doubling --- effective two frequency forcing
(now at $f/2$ and $f/4$) is restored and hexagons reappear.  Figure
\ref{pics}(e) shows that four different phases simultaneously exist in
the system; each phase present before the second period
doubling supports two separate phase domains of hexagons.  As Fig.\
\ref{phasd} indicates, the transition to $f/4$ hexagons is
non-hysteretic.


The layer becomes spatially and temporally disordered for $\Gamma
\approx 7.6$.  Disorder is introduced by circular regions
approximately three cells in diameter, which randomly appear in the
hexagonal patterns and then shrink and disappear over
the course of approximately $50/f$.  The circular regions are $\pi /
4$ out of phase with the domain that contains them and result from a
single flight of $t_{{\rm flt}} \approx 1/f$, instead of $t_{{
\rm flt}} \approx 2/f$ as for
the rest of the domain.  For $\Gamma < 7.8$ the four phase domains
associated with the hexagonal pattern remain intact.  However, for
larger $\Gamma$, hexagons and the boundaries between the stable
domains disappear; the layer consists of numerous small domains of
wave-like structures with short spatial and temporal correlation (see
Fig.\ \ref{pics}(f)).  The disordered state persists up to $\Gamma =
14$, the largest forcing examined.


The ordered patterns we observe in granular media have similarities
with those in vertically oscillated fluids (Faraday experiment)
\cite{edwards}, but an important difference is that period doubling, 
which is not observed in fluids, leads to domains with different phase
in the granular layer.  With this exception, it appears that the
factors determining pattern selection in the two media are similar
even though, because of free flights and collisions, the effective
forcing of the granular layer is different from the smooth forcing of
the fluid.  Dilation of the granular layer appears to be analogous to
the inverse of viscosity in fluids because in the granular system we
observe $f/2$ squares for large dilation and stripes for small
dilation, while in fluids squares are observed for small viscosity
(less than 0.7 cm\footnotemark[2]/s) and stripes for large viscosity
\cite{edwards}. In the Faraday experiment hexagons appear when the
container is externally driven simultaneously at both $f$ and $f/2$,
but only for certain ranges of the relative phase and amplitude of the
two drive signals; if the phase difference is too small, squares are
observed \cite{muller}.  Hexagons appear in the granular layer when
the intrinsic relative phase associated with the effective
two-frequency forcing is large.  In thinner particle layers the
transition to effective two frequency forcing occurs for smaller
$\Gamma$; consequently the relative phase between long and short
flights is smaller and we observe patterns of squares with the same
peak and cell structure as for hexagons.


The pattern formation phenomena we have described result from the
interaction of parametric waves and period doubling.  In addition to
the results reported here, we have performed experiments varying the
particle restitution coefficient (0.5--0.95), density (2.3--11.4 g
cm\footnotemark[-3]) and size (0.05--3 mm); the layer thickness (2--40
particles) and aspect ratio (2--100); and the pressure
($10^{-1}$--$10^{-6}$ Torr).  No qualitative changes in the patterns
or their associated bifurcations were observed.  Moreover, in contrast
with most other cooperative dynamic phenomena in granular systems, we
have found that transition thresholds and pattern organization become
better defined as the pressure is reduced and the ratio of container
size to pattern wavelength is increased; this indicates that
interstitial gas and sidewall forcing do not organize the grain
motion.  Future work examining basic transport properties such as
viscosity and momentum diffusivity in the robust patterns we have
found should have practical consequences.


The work of F. M. was supported by Fondecyt Grant No. 1941115.  The
work of P. B. U. and H. L. S. was supported by the Department of
Energy, Office of Basic Energy Sciences Grant
No. DE-FG03-93ER14312. The Texas-Chile collaboration was supported by
a cooperative program between Conicyt and the National Science
Foundation, Division of International Programs, Americas Program Grant
No. INT-9415709.


%
%
\dofloatfig\else
\begin{figure}
\smallskip
\caption{Patterns in a 1.2 mm deep layer at $f$ = 67 Hz: (a) $f/2$
stripes ($\Gamma$ = 3.3), (b) $f/2$ hexagons ($\Gamma$ = 4.0), (c)
flat with kink ($\Gamma$ = 5.8), (d) competing $f/4$ squares and
stripes ($\Gamma$ = 6.0), (e) $f/4$ hexagons ($\Gamma$ = 7.4), and (f)
disorder ($\Gamma$ = 8.5).}
\label{pics}
\end{figure}


\begin{figure} 
\caption{Stability
diagram showing transitions in a 1.2 mm deep layer.  The vertical
dashed lines indicate the frequencies above which only stripes appear
in the wave regimes.  Closed (open) square and circular symbols denote
the transitions to waves with increasing (decreasing) $\Gamma$.}
\label{phasd} 
\end{figure}


\begin{figure} 
\caption{Comparison of observations for a particle layer to
calculations for a model consisting of a 1D completely inelastic ball.
(a) Flight times for the layer (1.9 mm deep, $f = 67$ Hz), $\tau_{{\rm
expt}}$ ($\bigtriangleup$), and for a completely inelastic ball,
$\tau_{{\rm calc}}$ ($\circ$).  (b) The dimensionless layer collision
acceleration $\gamma / \gamma_{{\rm crit}}$ (the average value is
displayed where $\gamma$ is multivalued).  (c) Container acceleration
(collisions removed). (d) and (e) Impact acceleration for stripes
($\Gamma=3.3$) and hexagons ($\Gamma = 4.2$) respectively (vertical
arrows indicate successive collisions of regions with the same
relative phase).}
\label{ftime} 
\end{figure}

\fi


\ifmypprint\else
  \vfill\eject
  \epsfxsize=5.5 truein  \centerline{\epsffile{umbanhowar1.eps}}
  \vfill{\small Melo {\it et al.}, Fig.~1} \vfill\eject
  \epsfxsize=6 truein  \centerline{\epsffile{umbanhowar2.eps}}
  \vfill{\small Melo {\it et al.}, Fig.~2} \vfill\eject
  \epsfxsize=6 truein  \centerline{\epsffile{umbanhowar3.eps}}
  \vfill{\small Melo {\it et al.}, Fig.~3} \vfill\eject
\fi


\begin{table}
\caption{Patterns and their associated instabilities.}
\label{table1}
\begin{tabular}{llcc}  
{\bf Pattern} & {\bf Instability} & $\tau_{{\rm expt}}$ & $\gamma / 
\gamma_{{\rm crit}}$ \\ \tableline 
flat & none & $< 1$ & $<1$ \\
$f/2$ waves\tablenotemark[1] & parametric & $< 1$ & $> 1$ \\ 
$f/2$ hexagons\tablenotemark[2] & parametric \& period-2& 
$>$,$<1$ & $> 1$ \\
flat with kinks & period-2 & $> 1$ & $< 1$ \\ 
$f/4$ waves\tablenotemark[1] & parametric \& period-2& $>1$ & $> 1$ \\
$f/4$ hexagons\tablenotemark[2]& parametric \& period-4& $>,<2$& $> 1$ \\ 
\end{tabular}
\tablenotetext[1]{squares or stripes}
\tablenotetext[2]{effective two-frequency forcing}
\end{table}


\begin{references}
\bibitem{gamma} The form of $\gamma$ is not critical.  A better
approximation is $\gamma = v_c/[g t_{{\rm flt}} (1 - \frac{(1 - \epsilon)
g t_p}{2 v_c})]$, where $\epsilon$ is the coefficient of restitution
and $t_p$ is the time spent on the plate; when $t_p / v_c$
is small the expression in the text is recovered.
\end{references}
\end{document}